\documentclass[12pt]{article}
\usepackage{amsmath, amssymb}
\usepackage{mathrsfs}
\begin{document}
\date{}
\title{On the supersymmetry of the Dirac-Kepler problem plus a Coulomb-type scalar potential in $D+1$
dimensions and the generalized Lippmann-Johnson operator}

\author{D. Mart\'{\i}nez$^{a}$\footnote{{\it E-mail address:} dmartinezs77@yahoo.com.mx}, M. Salazar-Ram\'{\i}rez$^{b}$, \\ R. D. Mota$^{c}$ and V. D. Granados$^{b}$} \maketitle

\begin{minipage}{0.9\textwidth}

\small $^{a}$ Universidad Aut\'onoma de la Ciudad de M\'exico,
Plantel Cuautepec, Av. La Corona 320, Col. Loma la Palma,
Delegaci\'on
Gustavo A. Madero, 07160, M\'exico D. F., M\'exico.\\

\small $^{b}$ Escuela Superior de F\'{\i}sica y Matem\'aticas,
Instituto Polit\'ecnico Nacional,
Ed. 9, Unidad Profesional Adolfo L\'opez Mateos, 07738 M\'exico D F, M\'exico.\\

\small $^{c}$ Escuela Superior de 
Ingenier\'{\i}a  Mec\'anica y El\'ectrica, Unidad Culhuac\'an, Instituto Polit\'ecnico Nacional, Av. Santa Ana No. 1000,   Colonia San Francisco Culhuac\'an, Delegaci\'on Coyoac\'an, C. P. 04430, M\'exico D. F., Mexico.\\

\end{minipage}

\begin{abstract}
We study the Dirac-Kepler problem plus a Coulomb-type scalar potential by generalizing the  Lippmann-Johnson operator to $D$ spatial dimensions. From this operator, we construct the supersymmetric generators to obtain the energy spectrum for discrete excited eigenstates and the radial spinor for the SUSY ground state.
\end{abstract}

\maketitle

\section{Introduction}
The vector potential is introduced into the Dirac equation by minimal coupling, whereas the scalar potential is added to the mass term. Thus, it can be interpreted as a position-dependent mass term. The vector Coulomb potential can be derived from the exchange of massless photons between the nucleus and the leptons orbiting around it, the Coulomb-type potential can be created by exchange of massless scalar mesons \cite{GREINERR}.  The scalar potential has been of great importance in the relativistic quark model, it was employed for describing magnetic moments and to avoid the Klein paradox risen in the quarkonium confining potentials \cite{TUTIK}. Also, a relativistic scalar potential has been used as a model for the spatially dependent valence and conduction band edges of semiconductors near the $\Gamma$ and $L$ points in the Brillouin zone \cite{JUNKER,COOPER}.

The Dirac-Kepler problem in $D+1$ dimensions has been treated in
several ways: by power series \cite{HECHT}, radial SUSY QM
\cite{SUKUMAR}, supersymmetry generated by the Lippmann-Johnson
operator \cite{JORGENSEN} and an $su(1,1)$ approach \cite{NUESTRO}.
On the other hand, the Dirac radial equations with vector and scalar
Coulomb-type potentials were studied by series power \cite{GREINER},
SUSY QM \cite{CHINOS} and intertwining operators \cite{LEVIATAN}.

Joseph was the first who studied the Dirac equation with vector
potential in $D$ spatial dimensions \cite{JOSEP}. Also, the
Lippmann-Johnson operator was introduced to generate the
supersymmetric charges \cite{KATSURA}. With Coulomb-type scalar and
vector potentials,  the problem was solved in analytical way by
reducing the radial Dirac equations to the differential equations
satisfied by confluent hypergeometric functions \cite{TUTIK,DONG}. The
energy spectrum and the SUSY ground state of this problem were found
by an $su(1,1)$ algebraic approach \cite{MANUEL}. 
Notice that in order to the Coulomb potential obeys the Gauss' law, it must be of the form $-\alpha/r^{D-2}$. However, it has been shown that this potential leads to unstable orbits at the classical regime \cite{EREN, SWIEBACH} whereas for the non-relativistic quantum treatment for bound states there are not normalizable wave functions \cite{BRASIL}. These features have compelled to set the Coulomb potential in higher dimensions as the $-\alpha/r$ potential \cite{TUTIK, KATSURA,DONG, MANUEL}. In the present work we restrict the vector and scalar potentials to have this form. To our knowledge the Lippmann-Johnson operator in general dimensions with both potentials has not been reported and as a consequence, the supersymmetry generated from this constant of motion remains untreated

The purpose of this Letter is to construct the supersymmetry charges
for the Dirac-Kepler problem plus a Coulomb scalar potential from
the generalized Lippmann-Johnson operator in $D$ dimensions. The
fact that one of the supercharges annihilates the SUSY ground state
leads us to find the energy spectrum for discrete excited states and
to obtain the radial differential equations for the SUSY ground
state. By performing a similarity transformation to the radial
Lippmann-Johnson operator we find the SUSY ground state. Also, we
show that the radial part of the Lippmann-Johnson operator is
reduced to that reported in \cite{LEVIATAN} for the
three-dimensional space and obtained by intertwining considerations.

\section{The relativistic Dirac equation in $D+1$ dimensions and the generalized Lippmann-Johnson operator}
The Dirac equation in $D+1$ dimensions for a central field can be
written with  ($\hbar=c=1$) as \cite{GREINER}
\begin{gather}\label{Dir}
H\Psi\equiv\Big\lbrace\gamma^0\gamma^j p^j+\gamma^0\left(m+V_s\left(r\right)\right)+V_v\left(r\right)\Big\rbrace\Psi=i\frac{\partial\Psi}{\partial{t}},\hspace{2ex} j=1,2,...,D
\end{gather}
where summation over repeated index is assumed, $m$
is the mass of the particle, $V_s$ and $V_v$ are the spherically
symmetric scalar and vector potentials, respectively, and the Dirac
matrices in $D$ dimensions, $\gamma^j$, satisfy
$\lbrace\gamma^\mu,\gamma^\nu\rbrace=2\eta^{\mu\nu}$ with
\begin{equation}
\eta^{\mu\nu}=\begin{cases}
\delta^{\mu\nu} & \text{if $\mu=0$,}
\\
-\delta^{\mu\nu} &\text{if $\mu\neq 0$.}
\end{cases}
\end{equation}

In $D$ spatial dimensions, the orbital angular momentum operator
$L_{ab}$ and the total angular momentum $J_{ab}$ are defined as
\begin{equation}
L_{ab}=ix_a\partial_b-ix_b\partial_a
\end{equation}
and
\begin{equation}
J_{ab}=L_{ab}+\frac{i}{2}\gamma^a\gamma^b,
\end{equation}
respectively. In the case of spherically symmetric potentials, the total angular momentum
operator and the spin-orbit operator
\begin{equation}
K_D=-\gamma^0\Bigg\lbrace\frac{i}{2}\sum_{a\neq b}\gamma^a\gamma^bL_{ab}+\frac{1}{2}\left(D-1\right)\Bigg\rbrace
\end{equation}
commute with the Dirac Hamiltonian. For a given total angular
momentum $j$, the eigenvalues of $K_D$ are
$\kappa_D=\pm\left(j+1/2\right)$, where the minus sign is for
aligned spin $j=\ell+\frac{1}{2}$, and the plus sign is for
unaligned spin $\ell-\frac{1}{2}$ \cite{DONG}.

We find that if the scalar and vector potentials are given by
$V_s=\frac{\alpha_s}{r}$ and $V_v=\frac{\alpha_v}{r}$ then, the
matrix hermitian operator
\begin{equation}
B=-iK_D\gamma^{D+1}\left(H-\gamma^0m\right)+\gamma^{D+1}\gamma^0\gamma^i\frac{x^i}{r}\left(\alpha_v m+\alpha_s H\right)
\end{equation}
is a constant of motion (see appendix A), where the pseudoscalar
$\gamma^{D+1}$ is reduced to the matrix $\gamma^5$ in $(3+1)$
dimensions and satisfies
$\left(\gamma^{D+1}\right)^\dag=\gamma^{D+1}$,
$\left(\gamma^{D+1}\right)^2=1$ and $\lbrace\gamma^{D+1},
\gamma^\mu\rbrace=0$. Also, it can be shown that this operator
anticommutes with the Dirac operator $K_D$. Therefore, the operator
$B$ satisfies
\begin{equation}
B\Psi_{\kappa_D}=-b\Psi_{-\kappa_D},
\label{Beq}
\end{equation}
where $b$ is an undetermined constant. In fact, $B$ is the
generalization of the Lippmann-Johnson operator \cite{JOHNSON} and
is reduced to that given in \cite{LEVIATAN} for $D=3$, and to that
for $D$ dimensions in absence of the scalar potential $V_s$ reported
in \cite{KATSURA}.

In this way, for odd or even dimensions, we write the eigenstates of
equation (\ref{Dir}) as
\begin{equation}\label{spinor}
\Psi_{\kappa_D}=r^{-\frac{D-1}{2}}
\begin{pmatrix}
G_{\kappa_D}^{(1)}\left(r\right)\chi_{\kappa_D}^\mu\left(\Omega_D\right)\\
iG_{\kappa_D}^{(2)}\left(r\right)\chi_{-\kappa_D}^\mu\left(\Omega_D\right)
\end{pmatrix}e^{-iE t},
\end{equation}
being $G_{\kappa_D}^{(1)}$ and $G_{\kappa_D}^{(2)}(r)$ the radial
functions, and $\chi_{\kappa_D}^{\mu}\left(\Omega_D\right)$ the
hyperspherical harmonic functions coupled with the angular momentum
$j$ \cite{DONG}. We consider that
\begin{equation}
\gamma^{D+1}\gamma^0\gamma^i\frac{x^i}{r}\Psi_{\kappa_D}=-\Psi_{-\kappa_D},\label{gr}
\end{equation}
which is the generalization of the three-dimensional equation
$\left(\vec\sigma\cdot\hat r\right) \Psi_{\kappa}=-\Psi_{-\kappa}$
\cite{JORGENSEN,SAKURAI}. By defining the spinors
\begin{equation}
\Theta_{\kappa_D}=\begin{pmatrix}
-G_{\kappa_D}^{(2)}\chi_{\kappa_D} \\
iG_{\kappa_D}^{(1)}\chi_{-\kappa_D}
\end{pmatrix}e^{-iE t}
,\hspace{1ex}
\Phi_{\kappa_D}=\begin{pmatrix}
G_{\kappa_D}^{(1)}\chi_{\kappa_D} \\
iG_{\kappa_D}^{(2)}\chi_{-\kappa_D}
\end{pmatrix}e^{-iE t}
\label{spinors}\end{equation}
and from the results of Appendix B, the explicit form of the
operator $B$ acting on a general eigenstate of the Dirac Hamiltonian is
\begin{multline}
B\Psi_{\kappa_D}=-r^{-\frac{D-1}{2}}\Big\lbrace\left(\alpha_s\partial_r-\kappa_D V_v\right)\Theta_{\kappa_D}+\left(\kappa_D\left(\partial_r+\frac{\kappa_D}{r}\gamma^0\right)+m\alpha_v\right)\Phi_{\kappa_D}\\
+\alpha_s\gamma^0\left(m+V_s+\gamma^0 V_v\right)\Phi_{\kappa_D}\Big\rbrace.\label{Bpsi}
\end{multline}

\section{SUSY QM and the energy spectrum}
Based on references \cite{JORGENSEN,KATSURA,TANGERMAN}, we define
the supersymmetric generator
\begin{equation}
Q=\begin{pmatrix} 0 & 0\\
B & 0
\end{pmatrix},
\end{equation}
which satisfies $\lbrace Q,Q\rbrace=0$ and
\begin{equation}
{\cal H}\equiv\lbrace
Q,Q^\dag\rbrace=\begin{pmatrix} B^2 & 0\\
0 & B^2
\end{pmatrix},\label{Hsusy}
\end{equation}
with $\cal H$ the supersymmetric Hamiltonian.

In order to obtain the energy spectrum for the Dirac Hamiltonian
$H$, we consider the results given in Appendix C, from which
\begin{equation}
B^2=\left(\alpha_v m+\alpha_s H\right)^2+K_D^2\left(H^2-m^2\right).
\end{equation}
By considering equation (\ref{Beq}), we have the eigenvalue equation
$B^2\Psi_{\kappa_D}=b^2\Psi_{\kappa_D}$. Thus
\begin{equation}
b^2=\left(\alpha_v m+\alpha_s
E\right)^2+\kappa_D^2\left(E^2-m^2\right).\label{bb}
\end{equation}
Since the supersymmetric ground state, $\Psi_{SUSY}^0$ must satisfy
the condition
\begin{equation}
{\cal H}\Psi_{SUSY}^0=0,\label{susy}
\end{equation}
it follows that the ground state energy eigenvalue, $E_0$, is
obtained from equation (\ref{bb}) by setting $b=0$. In this way
\begin{equation}
E_0=m\Bigg\lbrace-\frac{\alpha_s\alpha_v}{\alpha_v^2+\gamma^2}\pm\sqrt{\left(\frac{\alpha_s\alpha_v}{\alpha_v^2+\gamma^2}\right)^2-\left(\frac{\alpha_s^2-\gamma^2}{\alpha_v^2+\gamma^2}\right)}\Bigg\rbrace\label{E0},
\end{equation}
where $\gamma^2=\kappa_D^2+\alpha_s^2-\alpha_v^2$. For the excited
states of the Hamiltonian $H$, we perform the
change $\gamma\rightarrow\gamma+n$, where $n=0,1,2,3...$ is the
radial quantum number. Then
\begin{align}
\frac{E_n}{m}=-\frac{\alpha_s\alpha_v}{\alpha_v^2+(\gamma+n)^2}\pm\sqrt{\left(\frac{\alpha_s\alpha_v}{\alpha_v^2+(\gamma+n)^2}\right)^2-\left(\frac{\alpha_s^2-(\gamma+n)^2}{\alpha_v^2+(\gamma+n)^2}\right)}\label{En},
\end{align}
which is in accordance to that obtained from an analytical
\cite{GREINER,DONG} or $su(1,1)$ algebraic approach \cite{MANUEL}.

The eigenstates $\Psi_{\kappa_D}$ and $\Psi_{-\kappa_D}$ are
transformed into each other by the operator $B$ (equation
(\ref{Beq})) and both are eigenfuntions of the operator $B^2$ with
the same eigenvalue. Therefore, the supersymmetric eigenstates can
be written as
\begin{equation}
\Psi_{SUSY}=\begin{pmatrix} \Psi_{\kappa_D}\\
\Psi_{-\kappa_D}\end{pmatrix}.\label{susystate}
\end{equation}

Considering equations (\ref{Hsusy}) and (\ref{susy}), the components
of the sypersymmetric ground state must satisfy
\begin{equation}
B\Psi_{0\;\pm\kappa_D}=0. \label{B=0}
\end{equation}
In order to solve this equation for $+\kappa_D$ (the solution for
the other sign can be obtained by equation (\ref{Beq})) we consider
the expression (\ref{Bpsi}), from which
\begin{gather}
B\Psi_{\kappa_D}=-r^{-\frac{D-1}{2}}e^{-iE t}\Bigg\lbrace\left(\alpha_s\partial_r-\kappa_D V_v\right)\begin{pmatrix}
-G_{\kappa_D}^{(2)}\chi_{\kappa_D} \\
iG_{\kappa_D}^{(1)}\chi_{-\kappa_D}
\end{pmatrix}\nonumber\\
+\left(\kappa_D\partial_r+m\alpha_v+\alpha_s V_v\right)\begin{pmatrix}
G_{\kappa_D}^{(1)}\chi_{\kappa_D} \\
iG_{\kappa_D}^{(2)}\chi_{-\kappa_D}
\end{pmatrix}\nonumber\\
+\left(\frac{\kappa_D^2}{r}+\alpha_s\left(m+V_s\right)\right)\begin{pmatrix}
G_{\kappa_D}^{(1)}\chi_{\kappa_D} \\
-iG_{\kappa_D}^{(2)}\chi_{-\kappa_D}
\end{pmatrix}\Bigg\rbrace.\label{Bpsi1}
\end{gather}

Thus, the radial components of the SUSY
ground state ($G^{(1)}_{0\;\kappa_D}$ and $G^{(2)}_{0\;\kappa_D}$) satisfy
\begin{equation}
L_D\begin{pmatrix}
G^{(1)}_{0\;\kappa_D}\\
G^{(2)}_{0\;\kappa_D}
\end{pmatrix}=\begin{pmatrix}
0\\0
\end{pmatrix},
\end{equation}
where
\begin{equation}
L_D\equiv
\begin{pmatrix}
\frac{d}{dr}+\frac{\epsilon_+}{r}+\frac{m\alpha_+}{\kappa_D} & -\frac{\alpha_s}{\kappa_D}\frac{d}{dr}+\frac{\alpha_v}{r}\\
\frac{\alpha_s}{\kappa_D}\frac{d}{dr}-\frac{\alpha_v}{r} & \frac{d}{dr}-\frac{\epsilon_-}{r}-\frac{m\alpha_-}{\kappa_D}
\end{pmatrix},\label{L}
\end{equation}
$\epsilon_{\pm}=\kappa_D+\alpha_s\alpha_{\pm}/\kappa_D$ and
$\alpha_{\pm}=\left(\alpha_s\pm\alpha_v\right)$. The matrix operator
$L_D$, obtained by means of SUSY QM, is reduced to the
three-dimensional operator $L$ reported in \cite{LEVIATAN}, which
has been constructed by imposing an intertwining relation between
the corresponding radial Dirac Hamiltonian and $L$.

In order to find the expression for the radial components of the
SUSY ground state, we define
\begin{equation}
\begin{pmatrix}
\tilde G_{0\;\kappa_D}^{(1)}\\
\tilde G_{0\;\kappa_D}^{(2)}
\end{pmatrix}=
\begin{pmatrix}
1 & -\frac{\alpha_s}{\kappa_D}\\
\frac{\alpha_s}{\kappa_D} & 1
\end{pmatrix}\begin{pmatrix}
G_{0\;\kappa_D}^{(1)}\\
G_{0\;\kappa_D}^{(2)}
\end{pmatrix}
\label{d1}.
\end{equation}
Therefore, equation (\ref{L}) is rewritten as
\begin{equation}
\Bigg\lbrace\frac{d}{dr}+\frac{1}{r}\begin{pmatrix}
\kappa_D & \alpha_+\\
\alpha_- & -\kappa_D
\end{pmatrix}\Bigg\rbrace\begin{pmatrix}
\tilde{G}_{0\;\kappa_D}^{(1)}\\
\tilde{G}_{0\;\kappa_D}^{(2)}
\end{pmatrix}=
-\frac{m\kappa_D}{\kappa_D^2+\alpha_s^2}\begin{pmatrix}
\alpha_+ & \frac{\alpha_s\alpha_+}{\kappa_D}\\
\frac{\alpha_s\alpha_-}{\kappa_D} & -\alpha_-
\end{pmatrix}\begin{pmatrix}
\tilde{G}_{0\;\kappa_D}^{(1)}\\
\tilde{G}_{0\;\kappa_D}^{(2)}
\end{pmatrix},\label{L1}
\end{equation}
which can be easily solved by diagonalizing the matrix of the factor
$1/r$. For this purpose, we perform the transformation
\begin{align}
\begin{pmatrix}
\tilde F_{0\;\kappa_D}^{(1)}\\
\tilde F_{0\;\kappa_D}^{(2)}
\end{pmatrix}& =
\begin{pmatrix}
\kappa_D+\gamma & \alpha_+\\
-\alpha_- & \kappa_D+\gamma
\end{pmatrix}\begin{pmatrix}
\tilde{G}_{0\;\kappa_D}^{(1)}\\
\tilde{G}_{0\;\kappa_D}^{(2)}
\end{pmatrix}\label{d2}.
\end{align}
Thus, from equation (\ref{L1}), we obtain
\begin{equation}
\Bigg\lbrace\frac{d}{dr}+\frac{1}{r}\begin{pmatrix}
\gamma & 0\\
0 & -\gamma\\
\end{pmatrix}
\Bigg\rbrace\begin{pmatrix}
\tilde F_{0\;\kappa_D}^{(1)}\\
\tilde F_{0\;\kappa_D}^{(2)}
\end{pmatrix}=
-\frac{m}{\kappa_D^2+\alpha_s^2}
\begin{pmatrix}
\alpha_v\kappa_D+\alpha_s\gamma & 0\\
0 & \alpha_v\kappa_D-\alpha_s\gamma \\
\end{pmatrix}\begin{pmatrix}
\tilde F_{0\;\kappa_D}^{(1)}\\
\tilde F_{0\;\kappa_D}^{(2)}
\end{pmatrix}.\label{L2}
\end{equation}

The unnormalized solutions for these differential equations are given by
\begin{align}
\tilde{F}^{(1)}_{0\;\kappa_D}& =
r^{-\gamma}\exp\left(-\frac{m}{\kappa_D^2+\alpha_s^2}\left(\alpha_v\kappa_D+\alpha_s\gamma\right)r\right),\\
\tilde{F}^{(2)}_{0\;\kappa_D}& =
r^{\gamma}\exp\left(-\frac{m}{\kappa_D^2+\alpha_s^2}\left(\alpha_v\kappa_D-\alpha_s\gamma\right)r\right).
\end{align}
Since $\tilde{F}^{(1)}_{0\;\kappa_D}$ diverges at $r=0$, it is not a
physically acceptable solution. Hence, the radial spinor for the
supersymmetric ground state is
\begin{equation}
\psi_{SUSY}^0\equiv\begin{pmatrix}
0 \\
r^{\gamma}\exp\left(-\frac{m}{\kappa_D^2+\alpha_s^2}\left(\alpha_v\kappa_D-\alpha_s\gamma\right)r\right)
\end{pmatrix}.
\end{equation}
For the case $\alpha_s=0$, $\psi_{SUSY}^0$ is a normalizable
solution only for $\kappa_D<0$ which is in accordance to the results
reported in \cite{JORGENSEN} for $3+1$ dimensions.

Having obtained the SUSY ground state, the explicit form of the
eigenfunctions corresponding to higher SUSY energy levels should be
determined by solving the eigenvalue equation ${\cal
H}\Psi_{SUSY}=b^2\Psi_{SUSY}$ which is equivalent to find the
solutions of the equation $B^2\Psi_{\kappa_D}=b^2\Psi_{\kappa_D}$.
Nevertheless, this problem is much more complicated to solve than
the original Dirac eigenvalue equation.

It is worth noting that the excited 
energy levels for the Hamiltonian $H$ could be obtained in a different way by applying the 
Sukumar \cite{SUKUMAR} or the Thaller \cite{THALLER} approaches for the shape-invariant uncoupled second order Dirac
equations. However, in the present Letter we used the generalization of
the Lippmann-Johnson operator in $D$ dimensions and SUSY QM to obtain the energy spectrum for this problem.

\section{Concluding remarks}
We treated the Dirac-Kepler problem plus Coulomb-type scalar
potential by generalizing the Lippmann-Johnson operator in general
dimensions. This operator allowed us to construct the supersymmetric
charges from which we found the ground state energy spectrum and the
spectrum energy for discrete excited states. The action of the
Lippmann-Johnson operator on the ground state of the Dirac equation
leads to the radial operator $L_D$, equation (\ref{L}), which
generalizes to $D$-dimensions that reported in \cite{LEVIATAN} for
the three-dimensional space. The ground state obtained in this work
is in full agreement with those reported in \cite{MANUEL} and it
reduces to that in three dimensions reported in \cite{CHINOS}. With the generalized 
Lippmann-Johnson operator reported in  the present Letter, we can construct an
$SO(4)$ symmetry treatment, similar to that given for the Kepler-Coulomb problem
in three spatial dimensions  \cite{CHEN}, which is 
a work in progress. 

\appendix
\numberwithin{equation}{section}
\section{Calculation of $[B,H]=0$}
For an arbitrary radial function $f(r)$, we show that $\left[K_D,f(r)\right]=0$. Moreover, we find that
\begin{align}
\left[H,\gamma^0\right]& =-2\gamma^ip^i,\\
\left[H,\gamma^{D+1}\right]& =-2\gamma^{D+1}\gamma^0\left(m+V_s\right),\\
\left[H,\gamma^{D+1}\gamma^0\gamma^a\frac{x^a}{r}\right]& =\frac{2i}{r}\gamma^{D+1}\gamma^0 K_D.
\end{align}
Hence,
\begin{align}
\left[B,H\right]& =-iK_D\left[H,\gamma^{D+1}\right]\left(H-\gamma^0 m\right)\nonumber\\
& +imK_D\gamma^{D+1}\left[H,\gamma^0\right]\nonumber\\
& +\left[H,\gamma^{D+1}\gamma^0\gamma^a\frac{x^a}{r}\right]\left(\alpha_v m+\alpha_s H\right)\nonumber\\
& =-2iK_D\gamma^{D+1}\gamma^0\lbrace\left(m+V_s\right)\left(H-\gamma^0 m\right)\nonumber\\
& -m\gamma^0\gamma^ap^a-\left(V_vm+V_sH\right)\rbrace\nonumber\\
& =0.
\end{align}

\section{Calculation of $B\Psi_{\kappa_D}$}
Considering the properties of the operator $K_D$, the algebra
satisfied by the matrices $\gamma^i$ and from equations (\ref{gr})
and (\ref{spinors}), we obtain
\begin{align}
-iK_D& \gamma^{D+1}\left(H-\gamma^0m\right)\Psi_{\kappa_D}\nonumber\\
& =i\Bigg\lbrace\gamma^{D+1}\gamma^0\gamma^a\frac{x^a}{r}\left[\frac{x^b}{r}p^b-\frac{i}{r}\left(\gamma^0K_D+\frac{D-1}{2}\right)\right]\nonumber\\
& +\left(V_v-\gamma^0V_s\right)\gamma^{D+1}\Bigg\rbrace K_D\Psi_{\kappa_D}\nonumber\\
& =\kappa_Dr^{-\frac{D-1}{2}}\Big\lbrace-\left(\partial_r+\kappa_D\gamma^0\right)\Phi_{\kappa_D}\nonumber\\
& +\left(V_v-\gamma^0V_s\right)\Theta_{\kappa_D}\Big\rbrace
\end{align}
and
\begin{align}
\gamma^{D+1}& \gamma^0\gamma^i\frac{x^i}{r}\left(\alpha_v m+\alpha_s H\right)\Psi_{\kappa_D}\nonumber\\
& =-r^{-\frac{D-1}{2}}\Big\lbrace \alpha_s\left(\partial_r-\frac{\kappa_D}{r}\gamma^0\right)\Theta_{\kappa_D}\nonumber\\
& +\left[\alpha_s\gamma^0\left(m+V_s+\gamma^0
V_v\right)+m\alpha_v\right]\Phi_{\kappa_D} \Big\rbrace.
\end{align}
Thus, the explicit form of the Lippmann-Johnson operator $B$ acting
on an eigenstate $\Psi_{\kappa_D}$ of the Hamiltonian $H$ is
\begin{multline}
B\Psi_{\kappa_D}=-r^{-\frac{D-1}{2}}\Big\lbrace\left(\alpha_s\partial_r-\kappa_D V_v\right)\Theta_{\kappa_D}+\left(\kappa_D\left(\partial_r+\frac{\kappa_D}{r}\gamma^0\right)+m\alpha_v\right)\Phi_{\kappa_D}\\
+\alpha_s\gamma^0\left(m+V_s+\gamma^0 V_v\right)\Phi_{\kappa_D}\Big\rbrace.\label{Bpsi}
\end{multline}

\section{Calculation of $B^2$}
With the definitions $A_1=H-\gamma^0m$ and $A_2=\alpha_v m+\alpha_s
H$, we find the following commutation relations
\begin{align}
\left[A_1,A_2\right]&=-2m\alpha_S\gamma^ip^i,\\
\left[A_1,\gamma^{D+1}\right]&=-2\gamma^{D+1}\gamma^0V_s,\\
\left[A_2,\gamma^{D+1}\right]&=-2\alpha_s\gamma^{D+1}\gamma^0\left(m+V_s\right),\\
\left[A_1,\gamma^0\gamma^i\frac{x^i}{r}\right]&=\frac{2i\gamma^0}{r}K_D+2V_s\gamma^i\frac{x^i}{r},\\
\left[A_2,\gamma^0\gamma^i\frac{x^i}{r}\right]&=2iV_s\gamma^0K_D+2\alpha_s\left(m+V_s\right)\gamma^i\frac{x^i}{r}.
\end{align}
Since $\left[K_D,\gamma^0\gamma^i\frac{x^i}{r}\right]=0$ and
$\lbrace K_D,\gamma^{D+1}\rbrace=0$, and because of the anti-hermiticity of $\gamma^j$ ($j=1,2,...,D$), we finally obtain
\begin{align}
B^2& =B^{\dag}B=A_1^2K_D^2+A_2^2-iA_1\gamma^0\gamma^i\frac{x^i}{r}A_2K_D\nonumber\\
&+iA_2\gamma^0\gamma^i\frac{x^i}{r}A_1K_D\nonumber\\
&=A_1^2K_D^2+A_2^2+2m\left(V_v+\gamma^0V_s\right)\gamma^0K_D^2\nonumber\\
&=\left(H^2-m^2\right)K_D^2+A_2^2\nonumber\\
&=\left(\alpha_v m+\alpha_s H\right)^2+K_D^2\left(H^2-m^2\right).
\end{align}

\section*{Acknowledgments}
This work was partially supported by SNI-M\'exico, COFAA-IPN,
EDI-IPN, SIP-IPN project number $20110127$ , and ADI-UACM project
number 7DA2023001.


\begin{thebibliography}{99}
\bibitem{GREINERR} Greiner W., {\it Relativistic Quantum Mechanics} (Springer-Verlag, Berlin) 2000.
\bibitem{TUTIK} Tutik, R. S. {\it J. Phys. A: Math. and Gen.}, {\bf25} (1992) L413.
\bibitem{JUNKER} Junker G., {\it Supersymmetric Methods in Quantum and Statistical Physics} (Springer-Verlag, New York) 1996.
\bibitem{COOPER} Cooper F., Khare A.  and  Sukhatme U., {\it Supersymmetry in Quantum Mechanics} (World-Scientific, Singapure) 2001.
\bibitem{HECHT} Hecht K.T., {\it Quantum Mechanics} (Springer-Verlag, New York) 2000.
\bibitem{SUKUMAR} Sukumar C.V., {\it J. Phys. A: Math. Gen.}, {\bf18} (1985) L697.
\bibitem{JORGENSEN} Dahl J.P. and Jorgensen T., {\it Int. J. Quantum Chem.}, {\bf53} (1995) 161.
\bibitem{NUESTRO} Salazar-Ram\'irez M. et. al., {\it J. Phys. A: Math. Theor.}, {\bf43} (2010) 445203.
\bibitem{GREINER} Greiner, W., M\"uller B. and Rafelski J., {\it Quantum Electrodymanics of Strong Fields} (Springer-Verlag, Berlin) 1985.
\bibitem{CHINOS} Guo-Xing J. and Zhong-Zhou R, {\it Commun. Theor. Phys.}, {\bf49} (2008) 319.
\bibitem{LEVIATAN} Leviatan, A., {\it Phys. Rev. Lett.}, {\bf 92} (2004) 20.
\bibitem{JOSEP}A. Joseph, {\it Rev. Mod. Phys.}, {\bf39} (1967) 829.
\bibitem{KATSURA} Katsura H. and Aoki H., {\it J. Math. Phys.}, {\bf 47} (2006) 032301.
\bibitem{DONG} Dong S.H., Sung G.H. and Popov D., {\it J. Math. Phys.}, {\bf 44} (2003) 4467.
\bibitem{MANUEL} Salazar-Ram\'irez M. et. al., {\it Eur. Phys. Lett.}, {\bf 95} (2011) 60002.
\bibitem{EREN} Ehrenfest P., {\it Ann. Phys.}, {\bf  61}, (1920) 440.
\bibitem{SWIEBACH}Swiebach B., {\it A First Course in String Theory}, (Cambridge University Press, UK) 2009.
\bibitem{BRASIL}Braga N. R. F. and D'Andrea R., {\it arXiv:quant-ph/0511078v1} 
\bibitem{JOHNSON} Johnson M.H. and Lippmann B.A., {\it Phys. Rev.} {\bf 78} (1950) 329A.
\bibitem{TANGERMAN} Tangerman D.R. and Tgon J.A., {\it Phys. Rev. A}, {\bf 48} (1993) 1089.
\bibitem{SAKURAI} Sakurai J.J., {\it Advanced Quantum Mechanics} (Addison-Wesley, Reading MA) 1967.
\bibitem{THALLER} Thaller B., {\it Advanced Visual Quantum Mechanics} (Springer, USA) 2005.
\bibitem{CHEN} Chen J.L. Deng D.L. and Hu M.G., {\it Phys. Rev. A}, {\bf 77} (2008) 034102.
\end{thebibliography}
\end{document}